\documentstyle[epsf]{lamuphys}
\makeatletter
\let\chapter\hid@chapter
\makeatother

\begin{document}
\pagenumbering{arabic}
\title{Accretion and Jet Power}

\author{Beverley J.\,Wills}

\institute{Department of Astronomy, University of Texas at Austin, Texas,
78712}

\maketitle

\begin{abstract}
In the first of a series of three lectures we discuss ways of measuring the
power
available to feed the jets in powerful FR\,II radio sources.  For unobscured
radio-loud
QSOs we present evidence that this power is directly related to the UV-optical
luminosity, or probably more accurately, to the power radiated through
processes of
accretion in a strong gravitational potential.  It has been suggested on
theoretical grounds
that powerful radio
jets are a necessary component of the central engine.  It then follows, from
the similarity
of the optical-UV power output, spectral energy distribution, and emission-line
spectra of
radio-loud
and radio-quiet QSOs, that radio-quiet QSOs have the same power available to
feed jets as
do radio-loud QSOs.  This then leaves us with the puzzle of why we do not see
the
powerful jets in radio-quiet QSOs.
\end{abstract}

\section{Introduction}

The enormous luminosity of QSOs\footnote
{`QSO' refers to all luminous AGN (L$\ga 10^{11}$\,L$_{\sun}$, H$_0 =
 100$\, km s$^{-1}$\,Mpc$^{-1}$, and QSOs are called `radio-loud' if
F$_{\rm5GHz}$/F$_{4400\,\AA} \ga 10$, where F is the rest frame flux-density in
mJy;
such strong radio emission is assumed to indicate powerful radio jets.}
 is believed to arise from accretion of gas through a
large gravitational potential, over distance scales of $\sim 0.001$\ pc, giving
rise
to the characteristic optical-ultraviolet-X-ray spectral energy distributions
with T $\sim 10^4 {\rm -} 10^5$\,K (the Big Blue Bump).  Material is carried
inwards in an
accretion disk.  Conservation of angular momentum requires the transport of
angular
momentum and energy outwards.  One popular idea for achieving this
(K\"onigl \& Kartje 1995, but see Kundt \& Gopal-Krishna 1980) is that the
disk, consisting of ionized plasma, is threaded by magnetic field lines that
are twisted
by the disk's rotation.  Plasma flows out along the fields lines that then
leave the disk
perpendicular to the surface, allowing the plasma to be accelerated into
collimated,
relativistic outflows along the disk axis.  Thus jets regulate the transport of
angular
momentum and energy; in this hypothesis they are a necessary component of the
central engine,
enabling the accreting fuel to spiral inwards.

The optical-UV-X-ray spectral energy distributions are remarkably similar for
radio-loud and and radio-quiet QSOs (RLQs and RQQs).
So also are the strong, broad emission-line spectra
that arise in the surrounding high velocity gas of the broad line region (BLR),
and the
narrow-line spectra that arise from lower-velocity gas of the narrow line
region (NLR) at
greater nuclear distances ($\sim 1$\, pc -- Kpc); both the BLR and NLR are
ionized by the
0.01 -- 1 keV photons from this central continuum source.  These similarities
strongly
suggest the same type of central engine in RLQs and RQQs.

Even if we do not know the exact processes by which jets are formed, there are
two
empirical
approaches that suggest that the power that feeds the jets is directly
related to the power radiated by the central engine in the accretion process.
We
discuss these in \S2 and \S3.

\section{Jet Power, and Luminosity Relationships}

Two relations together show that jet power and accretion power (assumed to be
represented by power radiated by QSO's Big Blue Bump)
 are strongly coupled --
one between emission-line and optical-continuum luminosity, and the other
between
luminosity of emission lines and extended radio emission.

Yee (1980) and Shuder (1981) have shown that emission-line luminosities (e.g.,
H$\alpha$)
are closely proportional to the luminosity of the non-stellar, featureless
continuum at
4800 \AA\ rest wavelength, over a range of more than $10^5$\ in luminosity, for
a
heterogeneous assortment of narrow and broad line Seyfert and radio galaxies,
RQQs
and RLQs.  They show that this relationship is consistent with photoionization
of
the emission-line regions by an extrapolation of the observed non-stellar
continuum.

Rawlings \& Saunders (1991) report a close proportionality between the total
kinetic
power of the Kpc to Mpc-scale jets and narrow-line luminosity, for a complete
sample of
3CR FR\,II radio galaxies.  Despite their careful, detailed calculations of
kinetic jet
power, the result is not very different if simply the radio luminosity is used
instead,
provided any beamed core emission is excluded.
The completeness of the sample is important, to demonstrate that
the proportionality is not simply induced by the bias towards selecting sources
of
increasing radio and optical luminosity at higher redshift.  Such an unbiased
sample is not
available for RLQs; however, if we do include radio and narrow emission-line
data for
FR\,I radio galaxies, and RLQs with z$ < 1$, this relation extends over a range
of
10$^4$\ in luminosity.

The first relation, between emission-line and continuum luminosity, shows that
the ionizing
continuum is closely related to the
observed optical continuum, and that both are measured at least roughly, by the
emission-line luminosity.  This is useful because, in orientation-dust Unified
Schemes,
FR\,II radio galaxies are RLQs with the central engine hidden or partially
hidden from
the observer by a dusty torus or
inner galaxy disk whose axis is parallel to the jet direction.  In this case,
even
though the RLQ continuum and broad line region may be obscured, much of the
extended
gas is still illuminated and so narrow-line luminosity can apparently be used
as a measure
of the ionizing continuum and hence the accretion power.
The second relation, between emission-line and extended radio luminosity,
therefore shows
that the kinetic power of the jet is directly
related to the EUV-optical luminosity -- the accretion power.
Moreover, Rawlings \& Saunders argue that the central engine channels a
significant
fraction of power into the jets compared with that radiated by accretion; this
high
efficiency implies a
massive, spinning object that both powers the jets and controls the accretion
rate.

Impressive as these relations are, the scatter in them is almost an order of
magnitude.
This scatter could be accounted for entirely by the uncertainties in
determining
jet power and by how well line luminosity measures the ionizing continuum.
Significant
scatter is certainly expected from variations in gas covering, optical depth to
ionizing
photons, dust reddening and obscuration.  In other words, the true relationship
between jet
power and accretion power may actually be much tighter.

\section{Radio Core-dominance Relationships}

It is conventional to use the core dominance, R, as a measure of the
Doppler-boosting of
radio synchrotron emission arising from relativistic flows at the base of the
jet.
R is defined as the ratio of flux-density of the compact radio source
coincident with the
QSO nucleus, to the (assumed isotropic) flux-density in the extended radio
lobes,
measured in the QSO rest frame.  The nuclear flux density is typically measured
with
VLA resolutions of $\sim 1$\arcsec, and the nuclear source is usually
unresolved at
these resolutions.
If the luminosity in the extended lobes is a good measure of the power
available to feed
the nuclear jets, and the bulk velocities at the base of the jet are similar
from one
RLQ to another, then R should be an indicator of the angle $\phi$, of the jet
to the
line-of-sight.  On the simplest beaming hypotheses, R may range over factors of
$\sim 10^4$\,-- 10$^5$, for blazars with jets within a few degress of the line
of sight
(small $\phi$) to radio-galaxies with jets near the sky-plane ($\phi \sim
90$\degr).

Jets depend on synchrotron losses for their visibility, and there is strong
evidence that
their radio emission beyond several tens of Kpc from the nucleus depends
increasingly on
interactions with the
host galaxy or intergalactic environment (Bridle et al. 1994).  Their emission
may be beamed
even on scales of many Kpc.  Also, they represent jet power averaged over
millions of
years.  For these reasons extended radio luminosity may not be a good measure
of present
jet power, and so we looked for another way to normalize the beamed core
flux-density
(see Wills \& Brotherton 1995 for further discussion and references).
We found that the use of optical
luminosity for UV-excess FR\,II RLQs thought to have low reddening and
negligible
 synchrotron
contribution, significantly improved two relationships -- those of R with jet
angle and
R with width (FWHM) of the broad H$\beta$\ emission line.  The jet angle,
$\phi$\, is
determined completely independently from measurements of superluminal motion
and limits
on inverse Compton scattered X-ray flux.  These relationships are
compared in Figs. 1 and 2.
\begin{figure}
\plotfiddle{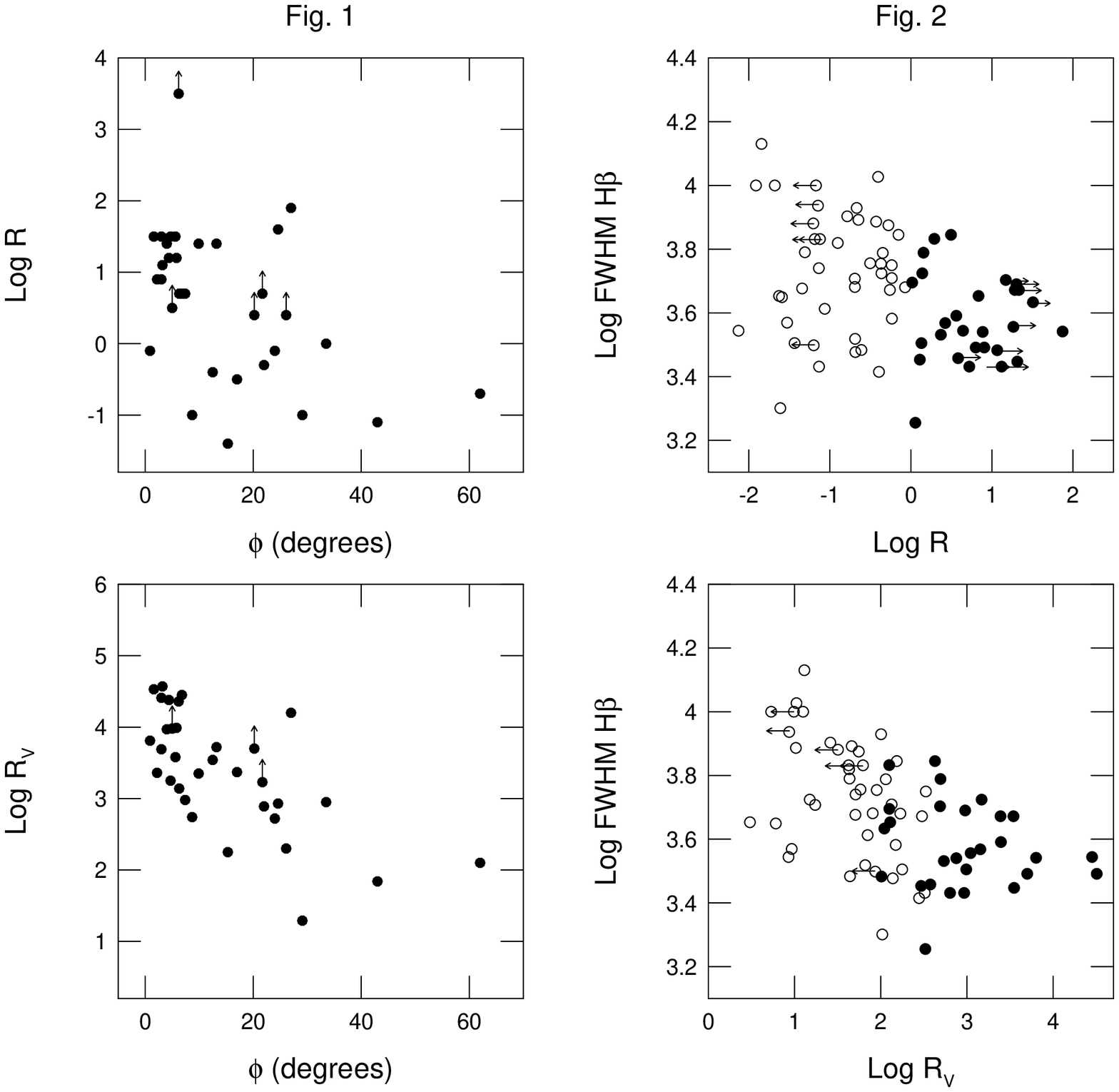}{12.5cm}{0.}{63}{63}{-197.5}{-70}
\caption{Radio core-dominance vs. $\phi$, the angle of the beam to the line of
sight.  The 2-tailed probability that these variables are unrelated decreased
from
$10^{-2}$\ to $10^{-5}$\ by using a rest-wavelength V-band luminosity to
normalize
core-dominance R.}
\caption{Radio core-dominance vs. the width of the broad H\,$\beta$\ line.
Here, the 2-tailed
probability that these variables are unrelated decreased from $4 \times
10^{-3}$\ to
less than $10^{-5}$\ by using R$_V$\ instead of R.\hfil\break
(Figures are adapted from Wills \& Brotherton 1995.)}
\end{figure}

\begin{figure}
\plotfiddle{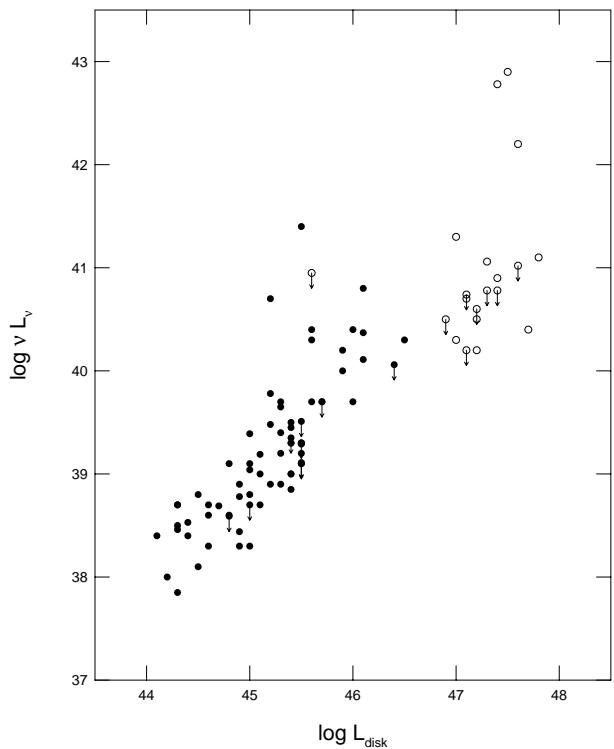}{10.5cm}{0.}{45.}{45.}{-140.}{-40.}
\caption{Total radio luminosity vs. disk (accretion) luminosity, for
radio-quiet
QSOs.  Data are for PG QSOs (the complete z $< 0.5$\ sample is shown as filled
circles,
higher redshift QSOs are shown as open circles).  The L$_{\rm disk}$\ values
are from
Falcke et al. (1995), with radio luminosities updated using Kellermann et al.
(1994).
Many previous upper limits on $\nu L_{\nu}$\ have become real measurements,
showing
even more convincingly the strong radio-UV luminosity relation.}
\end{figure}
\section{Radio Quiet QSOs}

Miller et al. (1992) and Falcke et al. (1995) show that there is a strong
emission line
vs. optical-UV continuum luminosity relationship for a complete,
optically-selected PG
sample, as also found for RLQs, showing a strong link between the observed and
ionizing (EUV) continua.  More importantly, despite the 100 times weaker radio
luminosity,
this complete sample of predominantly RQQs shows a closely proportional
relationship between total 5GHz luminosity and accretion luminosity (Miller et
al. 1993;
Falcke et al. 1995).  Fig. 3 shows this strong relationship from the paper by
Falcke et al.,
updated with more recent radio luminosities from Kellermann et al. (1994),
just for the PG QSOs.  The relationship also suggests
a narrow range of jet power to accretion power -- providing further support for
a real
dichotomy between RLQs and RQQs.

We suggest that investigations of this dichotomy via the distribution of radio
loudness for
QSOs should be made in the radio luminosity vs. optical-UV luminosity plane
rather
than, as has sometimes been done, in the single radio-luminosity dimension, or
even the
single dimension of the ratio of radio luminosity to optical-UV luminosity.
[Perhaps it
would be better to calculate radio jet power instead of radio luminosity
(Rawlings \& Saunders
1991), and a bolometric luminosity to represent accretion power, instead of
optical-UV
luminosity.]

\section{Summary and Discussion}

Previously-discovered relationships among emission line, optical continuum and
extended
radio luminosities indicate a proportionality between jet power and accretion
power in
high luminosity (FR\,II) sources.  The
improved correlations between core-dominance and jet angle, and between
core-dominance
and line width, when the beamed jet luminosity is normalized by the optical
continuum,
suggest an even stronger link on sub-parsec scales.

The similarity of the optical-to-EUV luminosities and spectra indicates the
same
accretion mechanism for the central engine of RLQs and RQQs
-- this is also suggested by the similar relationship between extended radio
luminosity and optical luminosity, albeit at very different radio luminosities.
These similarities, together with the conclusion that this relationship is a
strong
proportional one for QSOs with powerful radio jets, suggest that the same power
is potentially available to feed radio jets in RQQs.  It is therefore a real
puzzle that 90\% of QSOs are radio-quiet.

The conclusion that radio jet power depends directly on
ultraviolet-optical luminosity may seem to contradict the established belief
that there
is a very wide spread in extended radio luminosity for a given optical
luminosity.  It now
takes on new meaning to ask whether there is a real dichotomy in radio loudness
between
RLQs and RQQs.  If the central engines are identical, the differences
in (unbeamed) jet brightness must have an extrinsic cause and therefore be
dependent on
the environment beyond sub-parsec scales.
The question then becomes one of the distribution of environment conditions,
and whether
there is a discontinuity in the environmental conditions between RLQs and
RQQs.  An emission line approach to this will discussed in the next chapter.


\newpage
\begin{thebibliography}

\bibitem{}{}{}
Bridle\, A. H., Hough\, D. H., Lonsdale\, C. J., Burns\, J. O., Laing\, R. A.
(1994):
Deep VLA Imaging of Twelve Extended 3CR Quasars.
AJ {\bf 108}, 766--820
%
\bibitem{}{}{}
Emmering\, R. T., Blandford\, R. D., Shlosman\, I. (1992):
Magnetic Acceleration of Broad Emission Line Clouds in Active Galactic Nuclei.
ApJ {\bf 385}, 460--477
%
\bibitem{}{}{}
Falcke\, H., Malkan\, M. A. and Biermann\, P. L. (1995):
The jet-disk Symbiosis II. Interpreting the radio/UV Correlations in Quasars.
A\&A {\bf 298}, 375--394
%
\bibitem{}{}{}
Kellermann\, K. I., Sramek\, R. A., Schmidt\, M., Green\, R. F., Shaffer\, D.
B. (1994):
The Radio Structure of Radio-loud and Radio-quiet Quasars in the Palomar
Bright Quasar Survey.
AJ {\bf 108}, 1163--1177
%
\bibitem{}{}{}
K\"onigl\, A., Kartje\, J. F. (1994):
Disk-Driven Hydromagnetic Winds as a Key Ingredient of Active Galactic
Nuclei Unification Schemes.
ApJ {\bf 434}, 446--467
%
\bibitem{}{}{}
Kundt\, W., Gopal-Krishna (1980):
Extremely Relativistic Electron-positron Twin-jets form Extragalactic Radio
Sources.
Nature, 288, 149--150
%
\bibitem{}{}{}
Miller\, P., Rawlings\, S., Saunders\, R., Eales\, S. (1992):
A Spectrophotometric Study of BQS Quasars.
MNRAS {\bf 254} 93--110
%
\bibitem{}{}{}
Miller\, P. Rawlings\, S., Saunders\, R. {1993}:
The Radio and Optical Properties of the z $<$\,0.5 BQS Quasars.
MNRAS {\bf 263}, 425--460
%
\bibitem{}{}{}
Rawlings\, S., Saunders\, R. (1991):
Evidence for a Common Central Engine Mechanism in all Extragalactic
Radio Sources. Nat {\bf 349}, 138--140
%
\bibitem{}{}{}
Shuder\, J. M. (1981):
Emission-line--Continuum Correlations in Active Galactic Nuclei.
ApJ {\bf 244}, 12
%
\bibitem{}{}{}
Wills\, B. J., Brotherton\, M. S. (1995):
An Improved Measure of Quasar Orientation.
ApJ {\bf 448}, L81--L84
%
\bibitem{}{}{}
Yee\, H. K. C. (1980):
Optical continuum and emission-line luminosity of active galactic
nuclei and quasars.
ApJ {\bf 241} 894
%
\end{thebibliography}
\end{document}